\documentclass[conference]{IEEEtran}
\IEEEoverridecommandlockouts
\usepackage{graphicx}
\usepackage{subcaption}

\usepackage{cite}
\usepackage{float}
\usepackage{amsmath,amssymb,amsfonts}
\usepackage{graphicx}
\usepackage{textcomp}
\usepackage{xcolor}
\usepackage{algorithm} 
\usepackage{algpseudocode}

\usepackage[misc,geometry]{ifsym}

\usepackage{fancyhdr}

\def\BibTeX{{\rm B\kern-.05em{\sc i\kern-.025em b}\kern-.08em
    T\kern-.1667em\lower.7ex\hbox{E}\kern-.125emX}}

\begin{document}

% \title{Conference Paper Title*\\
\title{Dynamic Portfolio Optimization \textit{via} Augmented DDPG with Quantum Price Levels-Based Trading Strategy\\
\thanks{
*These authors contributed equally to this work.

\textsuperscript{\Letter}Corresponding author.}
}

% \author{
% \\
% \\
% \\
% \\
% \\
% \IEEEauthorblockN{Anonymous Authors}
% \\
% \\
% \\
% \\
% \\
% \\
% }

\author{\IEEEauthorblockN{1\textsuperscript{st} Runsheng Lin*}
\IEEEauthorblockA{
\textit{Department of Statistics and Data Science} \\
% \textit{Guangdong Provincial }\\
% \textit{Key Laboratory of Interdisciplinary}\\
% \textit{Research and Application for Data Science,}\\
\textit{BNU-HKBU United International College}\\
Zhuhai, China \\
p930026080@mail.uic.edu.cn}

\\
\IEEEauthorblockN{3\textsuperscript{rd} Mingze Ma}
\IEEEauthorblockA{
\textit{Department of Statistics and Data Science} \\
% \textit{Guangdong Provincial }\\
% \textit{Key Laboratory of Interdisciplinary}\\
% \textit{Research and Application for Data Science,}\\
\textit{BNU-HKBU United International College}\\
Zhuhai, China \\
p930006138@mail.uic.edu.cn}

\and
\IEEEauthorblockN{2\textsuperscript{nd} Zihan Xing*}
\IEEEauthorblockA{
\textit{Department of Statistics and Data Science} \\
% \textit{Guangdong Provincial }\\
% \textit{Key Laboratory of Interdisciplinary}\\
% \textit{Research and Application for Data Science,}\\
\textit{BNU-HKBU United International College}\\
Zhuhai, China \\
p930026137@mail.uic.edu.cn}

\\
\IEEEauthorblockN{4\textsuperscript{th} Raymond S.T. Lee\textsuperscript{\Letter}}
\IEEEauthorblockA{
\textit{Department of Computer Science} \\
% \textit{Guangdong Provincial }\\
% \textit{Key Laboratory of Interdisciplinary}\\
% \textit{Research and Application for Data Science,}\\
\textit{BNU-HKBU United International College}\\
Zhuhai, China \\
raymondshtlee@uic.edu.cn}
}

\maketitle

\begin{abstract}
 %Some researchers believe that supervised learning has the ability to predict asset prices and optimize portfolios, as revealed in several studies. However, due to the limitation that supervised learning itself is difficult to interact directly with the financial market, 
With the development of deep learning, Dynamic Portfolio Optimization (DPO) problem has received a lot of attention in recent years, not only in the field of finance but also in the field of deep learning.
Some advanced research in recent years has proposed the application of Deep Reinforcement Learning (DRL) to the DPO problem, which demonstrated to be more advantageous than supervised learning in solving the DPO problem. However, there are still certain unsolved issues: 1) 
DRL algorithms usually have the problems of slow learning speed and high sample complexity, which is especially problematic when dealing with complex financial data.
2) researchers use DRL simply for the purpose of obtaining high returns, but pay little attention to the problem of risk control and trading strategy, which will affect the stability of model returns. In order to address these issues, in this study we revamped the intrinsic structure of the model based on the Deep Deterministic Policy Gradient (DDPG) and proposed the Augmented DDPG model. Besides, we also proposed an innovative risk control strategy based on Quantum Price Levels (QPLs) derived from Quantum Finance Theory (QFT). Our experimental results revealed that our model has better profitability as well as risk control ability with less sample complexity in the DPO problem compared to the baseline models.
\end{abstract}


\section{Introduction}
% This document is a model and instructions for \LaTeX.
% Please observe the conference page limits. 
Dynamic Portfolio Optimization (DPO) is the process of adjusting asset allocation in a portfolio in a timely manner in order to ensure stable returns in a constantly changing market. Due to the complex, noisy, nonstationary and even chaotic nature of financial markets, traditional mathematical models are difficult to accomplish the task of DPO.

% With the development of Deep Learning (DL), many researchers have tried to use various time series forecasting models such as Recurrent Neural Networks (RNN) and Long-Short Term Memory (LSTM) networks to predict the price trend of financial products, to grasp the price trend and make appropriate trading actions in order to profit from the market. However, as revealed from the results of various studies, the performance of supervised learning in financial market forecasting is not so satisfactory. There is always a certain lag compared to ground truth data because of the noisiness of financial time series data. In addition, because supervised learning does not interact directly with the financial market, this resulted in an inherent weakness in many decision-making tasks such as DPO.
With the development of Deep Learning (DL), many researchers have tried to use various time series forecasting models such as Recurrent Neural Networks (RNN) and Long-Short Term Memory (LSTM) networks to predict the price trend of financial products, to grasp the price trend and make appropriate trading actions in order to profit from the market. Nevertheless, some inherent weaknesses in applying supervised learning to solve the DPO problem cannot be avoided. 
For example, supervised learning-based models require additional trading logic to translate into market behaviors of buying and selling, after predicting the price results. Meanwhile, the goal of its optimization while training, minimizing forecast errors, is different from the goal of maximizing the portfolio's returns in the DPO task. 
%Meanwhile, the performance of supervised learning-based trading models depends heavily on the accuracy of price prediction, which is a challenging task due to the noisiness of financial time series data.

However, Deep Reinforcement Learning (DRL), with the combination of Reinforcement Learning (RL) and DL, has been drawing the attention of researchers in recent years owing to its unique reward and punishment mechanisms, which makes it suitable for solving DPO problems in financial trading. Because through DRL, an accurate prediction output is not required; what the agent in the DRL algorithm pursues is just trading actions that can maximize the future profit based on DRL's unique reward and punishment mechanisms. With such characteristics, some researchers have tried to use DRL to accomplish the task of DPO. For example, value-based DRL methods such as DQN\cite{DQN} and policy-based methods such as DDPG\cite{https://doi.org/10.48550/arxiv.1509.02971} were applied to solve DPO problem and outperformed both traditional and advanced models\cite{gao2020application} \cite{zhang2021deep}, demonstrating that avoiding direct market prediction can indeed reduce the negative influence of noise in the market. 

Nevertheless, these works do not contribute to the solution of the slow learning speed of DRL algorithms while dealing with complex financial data. Also, risk control and trading strategy did not gain much attention from most of the researchers. Although the agent in DRL can acquire the knowledge of how to obtain high returns, its risk control and trading awareness are still not good enough to ensure the stability of the model’s performance. Therefore, our work aims to 1) make the training of the DRL model more efficient; 2) enhance the risk-control and stability of the DRL model. Based on such objectives, we proposed a new model based on DDPG and Policy Gradient (PG) known as Augmented DDPG. Inspired by \cite{9923640}, we merge the encoder network of Actor and Critic in DDPG, which increases the training speed of the model. %This is of interest for DRL models applied in DPO scenarios, where one can retrain the model at any time according to the changes in the market. 
In addition, we added a trading strategy based on Quantum Price Levels (QPLs), and the PG model learns this trading strategy to make the agent more risk-controlled and trade aware, resulting in better performance for our trading system.

% The remainder of this paper is organized as follows. Sec- tion II briefly discusses some related work. Section III in- troduces the proposed DRL day trading system. Section IV presents the experiments and results achieved with real market data from B3. Finally, Section V presents some conclusions and directions for future work.

\section{Related Work}

\subsection{Reinforcement Learning in Portfolio Optimization}

Taking the advance of DL, previous works \cite{zhang2020deep}\cite{wang2020portfolio} are conducted to solve the DPO problem based on the predictive power of deep neural networks for time series data, which shows good performance over traditional portfolio optimization theory-based methods. However, the shortcoming of supervised learning methods is rather obvious: trading with supervised learning-based methods necessitates additional effort besides forecasting, and its training goal is to minimize forecast error, which differs from the DPO task's goal of maximizing portfolio returns.

In recent years, a growing number of researchers have identified the limitation of supervised learning and tried to use DRL to optimize the solution. Basic DRL algorithms such as DQN \cite{DQN} and DDPG \cite{https://doi.org/10.48550/arxiv.1509.02971} are applied in DPO\cite{gao2020application} \cite{zhang2021deep} \cite{conegundes2020beating}, which demonstrate the potential of DRL models. In order to figure out how DRL can achieve better performance, a framework \cite{https://doi.org/10.48550/arxiv.1706.10059} based on DPG which applies different networks such as Convolutional Neural Networks (CNN), RNN and LSTM within the DPG is proposed, which tested on the cryptocurrency historical price data to compare the performance of various networks. Similarly, \cite{https://doi.org/10.48550/arxiv.1808.09940} explored the influences of different optimizers, target functions and learning rates in DRL models including DDPG, PPO and DPG, in DPO problems. With the expectation of better enhancing the horizon of DRL algorithms, \cite{ye2020reinforcement} collected financial news and used LSTM-based price trend prediction to augment the states of agents in DPG. Based on the basic DRL algorithms, \cite{yang2020deep} proposed an integrated strategy that integrates PPO, A2C, and DDPG to dynamically select models for portfolio allocation based on the performance of their Sharpe ratios. 

All these works have contributed significantly to the exploration of the DRL model in solving the DPO problem. However, these works have yet to make any progress in addressing the issue of DRL algorithms' low learning efficiency when dealing with complicated financial data. Moreover, risk control was not taken seriously by these efforts, and their risk control relied solely on the judgment of the agents in the DRL algorithm without giving them sufficient reference indicators.

\subsection{Quantum Finance Theory}

Quantum Finance Theory (QFT) introduced the intrinsic relationship between the financial market and the quantum harmonic model\cite{lee2020quantum}\cite{Meng_2015}. In QFT, the dynamics of financial products in the world financial market are financial particles with wave-particle duality. Financial particles have equilibrium states just like physical particles do. These particles will continue to exist in their current state in the absence of any external stimuli. However, if an outside stimulus has the power to excite a particle, it will migrate to a different energy level, either higher or lower. In the financial market, we refer to these levels as Quantum Price Levels (QPLs), which are adequate representations of the intrinsic energy of financial products\cite{lee2020quantum}. 

Because of the noisy nature of financial markets, traditional financial characteristics such as OPEN, HIGH, LOW, CLOSE, and even technical indicators such as the Moving Average (MA) and Relative Strength Index (RSI) are unable to describe the effective volatility of financial products accurately\cite{ATKINS2018120}. However, because QPLs model the energy level of the financial products themselves, different QPLs can effectively represent the volatility level of the financial product. 

Therefore, the QPLs-based indicator effectively captures the volatility of financial products\cite{qiu2021qf}. The principle to obtain such an indicator is to simulate the volatility of financial products using the Quantum Anharmonic Oscillator (QAHO) model and then solve it by Quantum Finance Schrödinger Equation (QFSE) to obtain the energy levels of financial products, which is QPLs. 

\section{Methodology}
\subsection{Problem Definition}
The purpose of this paper is to explore the effectiveness of Augmented DDPG in DPO tasks and the enhancement to model returns that result from a risk control module based on QPLs. DPO is the process by which an agent adjusts the allocation of assets in a portfolio in response to the changes in the financial market. This is done in order to ensure that the portfolio can maximize returns in the market with a certain level of risk control. Since we cannot accurately predict the future trends of financial markets, DPO tasks are suitable to be solved using model-free DRL models such as DDPG and PG. In such models, raw market data such as price returns will be input. The model will then optimize the parameters with the preset objective function as the optimization goal and output the optimal allocation of assets in a portfolio. 

To demonstrate that the DRL model has such capability, we will demonstrate the performance of our model in the form of back-testing. Therefore, we need to make the following assumptions to conduct back-testing,
\begin{itemize}
    \item[1)] \textbf{Adequate Liquidity}\\
    The term liquidity describes how quickly and easily a security or asset can be turned into cash without depreciating in value \cite{Liquidity}.\\
    \textbf{Assumption 1:} All market assets are liquid, and every transaction can be executed under the same conditions.
    \item [2)] \textbf{Zero Slippage}\\
    Slippage is the collective term for all instances in which a market player receives an alternative deal execution price from what was planned \cite{Slippage}.\\
    \textbf{Assumption 2:} All market assets are sufficiently liquid to allow for quick execution of trades at the most recent price when an order is made.
    \item[3)]\textbf{Zero Impact on the Market}\\
    The Law of supply and demand \cite{law} determines the price of assets in the market. Thus, any transaction alters the relative balance of the asset itself, which has an impact on the asset's price.\\
    \textbf{Assumption 3:} The trading agent's capital commitment is so minimal that it has no impact on the market.
\end{itemize}

\subsection{Quantum Price Levels}
The dynamics of financial products on global financial markets, such as forex, financial indices, commodities, etc., can be represented as Financial Particles (FPs) with wave-particle duality properties, according to Quantum Finance Theory (QFT)\cite{lee2020quantum}. If an external stimulus excites the particle, it will move to another energy level depending on the property of the stimulus. In the financial market, we refer to these levels as Quantum Price Levels (QPLs), which can also be thought of as support and resistance. 

To obtain QPLs, we need to solve the Quantum Finance Theory Schrödinger Equation (QFSE) by combining the Hamiltonian with traditional Schrödinger equation\cite{lee2020quantum}.

In QFT, Hamiltonian is defined as,
\begin{equation}
\widehat{H}=\frac{\hbar}{2 m} \frac{\partial}{\partial r^{2}}+V(r)
\end{equation}
where the Kinetic Energy (KE) is represented by $\frac{\hbar}{2m}\frac{\partial}{\partial r^2}$. $\hbar$ stands for the Plank constant and $m$ stands for the internal properties of the market, such as the market capital. $V\left(r\right)$ indicates the Potential Energy (PE). 

Therefore, the QFSE is formulated as,
\begin{equation}
\left[\frac{\hbar}{2 m} \frac{d^{2}}{d r^{2}}+\left(\frac{\gamma \eta \delta}{2} r^{2}-\frac{\gamma \eta v}{4} r^{4}\right)\right] \phi(r)=E \phi(r)
\end{equation}
where $E$ stands for the particle's energy levels, which for financial particles corresponds to the QPLs and $\phi\left(r\right)$ represents the QFSE wave-function, which is approximated by the probability density function of historical price return.

\subsection{Mathematical Formulation}
This section introduces some fundamental mathematical formulations in the DPO problem.
\begin{itemize}
    \item[1)] \textbf{Weight vector} (the percentage of each asset in the portfolio) on day $t$:\\
    \begin{equation}
\boldsymbol{w}_{\boldsymbol{t}}=\left(w_{0, t}, w_{1, t}, w_{2, t}, \ldots, w_{m, t}\right)^{T}
    \end{equation}
    where $m$ is the number of assets.
    \item [2)] \textbf{Execution price vector}
    	 (the execution price of each product) on day $t$:\\

    \begin{equation}
    \boldsymbol{v}_{\boldsymbol{t}}^E=\left(v_{1, t}^{\text {E}}, v_{2, t}^{\text {E }}, \ldots, v_{m, t}^{\text {E}}\right)^{T}
    \end{equation}
    where $m$ is the number of assets,
    % $\boldsymbol{v}_{\boldsymbol{t}}^E$ can be any number in the range $[\boldsymbol{v}_{\boldsymbol{t}}^{low},\boldsymbol{v}_{\boldsymbol{t}}^{high}]$
    $v_{i,t}^E$ can be any number in the range $[{v}_{{i,t}}^{low},{v}_{{i,t}}^{high}]$
    \item[3)]\textbf{Price return vector }	 (the return of each product) on day $t$:\\
    \begin{equation}  
    \boldsymbol{y}_{\boldsymbol{t}}=\left(1+r,\frac{\boldsymbol{v}_{\boldsymbol{t}}^E}{\boldsymbol{v}_{\boldsymbol{t-1}}^E}\right) 
    % &=\left(1+r, \frac{v_{1, t}^{\text {E}}}{v_{1, t-1}^{\text {close }}}, \frac{v_{2, t+1}^{\text {close }}}{v_{2, t}^{\text {close }}}, \ldots, \frac{v_{m, t+1}^{\text {close }}}{v_{m, t}^{\text {close }}}\right)^{T}
    \end{equation}
    where $r$ is the risk-free rate.
\end{itemize}
\subsection{DDPG Trading Agent}
% \subsubsection{\textbf{Observation} Space}
% For the trading agent, the only thing it can observe at time step t is the assets’ price, which is the price vector $V_t$,
% \begin{equation}
% \boldsymbol{o}_{\boldsymbol{t}}=\boldsymbol{v}_{\boldsymbol{t}}=\left(v_{1, t}^{\text {close }}, v_{2, t}^{\text {close }}, \ldots, v_{m, t}^{\text {close }}\right)^{T}
% \end{equation}

\subsubsection{\textbf{State} Space}
To obtain the state space at each time step, we processed the historical price and gained the log return matrix $r_{t-T:t}$, of windows size T. The reason for choosing log return instead of return is that by adding the log returns, we equivalently multiply the gross returns, and as a result, the networks can learn non-linear functions of the products of returns (i.e., asset-wise and cross-asset), which are the fundamental units of the covariances between assets \cite{https://doi.org/10.48550/arxiv.1909.09571}. In such a way, the relationship between different assets in the portfolio can be captured.\par
The formulation of the state is as follows,

\begin{equation}
\boldsymbol{s}_{\boldsymbol{t}}=\boldsymbol{r}_{\boldsymbol{t}-T: \boldsymbol{t}}=\left[\begin{array}{cccc}
r_{0, t-T} & r_{0, t-T+1} & \cdots & r_{0, t} \\
r_{1, t-T} & r_{1, t-T+1} & \cdots & r_{1, t} \\
\vdots & \vdots & \ddots & \vdots \\
r_{m, t-T} & r_{m, t-T+1} & \cdots & r_{m, t}
\end{array}\right]
\end{equation}
where  $r_{i,t-T:t}$ represents the log return of assets $i$ in the time interval of $[t-T,t]$ and $r_0 = \log(r)$.

\subsubsection{\textbf{Action} Space}
The solution to the DPO is to keep the agent updated on the asset allocation in the portfolio, that is, to determine the asset weight allocation for the next day $w_{t+1}$. Therefore, the action $a_t$ of the DRL agent should be the same as the weight vector $w_{t+1}$,
\begin{equation}
\boldsymbol{a}_{\boldsymbol{t}}=\boldsymbol{w}_{\boldsymbol{t + 1}}=\left(w_{0, t+1}, w_{1, t+1}, w_{2, t+1}, \ldots, w_{m, t+1}\right)^{T}
\end{equation}
% Thus, m-dimensional real space $R^m$ includes the action space A as a subset,

% \begin{equation}
%     \boldsymbol{a}_{t} \in A \subseteq R^{m}, \quad \forall t \geq 0 \quad \text { subject to } \sum_{i=0}^{M} a_{i, t}=1
% \end{equation}

\subsubsection{\textbf{Reward} with Transaction Costs \& Gini Bonus}
The definition of reward signals is usually the most challenging part of DRL because it determines what the learning goal of the agent is, which also determines the final performance of the agent. Therefore, our reward function should mimic as much as possible the real reward situation given by the market, while allowing the agent to have moderate risk management. Before introducing our definition of the reward signal, the following two terms should be illustrated.
\begin{itemize}
    \item[a)] \textbf{Transaction Costs}\\
    In a real trading market, transaction costs such as commission fees are an important factor affecting returns. Too many frequent transactions may make the transaction costs outweigh the benefits, resulting in an overall loss. Therefore, we should take transaction costs as an important consideration to simulate real market transactions.\par
    Before the market opens at time t, the weight vector of the portfolio is represented by $w_{t-1}$. Thus, as the value of each asset in the portfolio changes through time, the weight vector should be updated, 
    \begin{equation}
        \boldsymbol{w}_{t}^{\prime}=\frac{\boldsymbol{y}_{t-1} \odot \boldsymbol{w}_{t-1}}{\boldsymbol{y}_{t-1} \cdot \boldsymbol{w}_{t-1}}
    \end{equation}
    where $\odot $ is the element-wise multiplication.\par
    The weight change of assets should be defined as,
    \begin{equation}
        \Delta w_{i,t} = w_{i,t}-w_{i,t}^{\prime}
    \end{equation}
    Therefore, the transaction costs at time $t$ should be defined as,
    \begin{equation}
        \mu_{t}=C \sum_{i=1}^{m}\left|\Delta w_{i,t}\right|
    \end{equation}
    where $C$ is a constant that represents the commission rate.\par
    And then the cumulative return from time 0 to $T$ is formulated as follows,
    \begin{equation}
        R_{T}=\prod_{t=1}^{T}\left(1-\mu_{t}\right) \boldsymbol{w}_{\boldsymbol{t}} \cdot \boldsymbol{y}_{\boldsymbol{t}}
    \end{equation}
    \item[b)]\textbf{Gini Bonus}\\
    As mentioned above, apart from mimicking the transaction environment in the real market, we should also enable the agent to learn risk control management to some extent. In practice, we found that DRL models are very easy to “put all the eggs in one basket”, which means the portfolio almost invests in only one kind of asset. This situation would make the portfolio meaningless, while dramatically increasing the risk of the investment. \par
    Inspired by the Entropy Bonus \cite{wang2021commission}, we proposed the Gini Bonus, which gives a bonus to the action that diversifies the allocation of the portfolio. The Gini Bonus is formulated as,
    \begin{equation} \operatorname{Gini}\left(\boldsymbol{w_{t}}\right)=\sum_{i=1}^{m} w_{t, i}\left(1-w_{t, i}\right)=1-\sum_{i=1}^{m} w_{t, i}^{2}
    \end{equation}
\end{itemize}
Therefore, combining the above two terms, we defined the reward function of DDPG as,
    \begin{equation}
        reward_{t}=\log \left(\left(1-\mu_{t}\right) \boldsymbol{w_{t}} \cdot \boldsymbol{y_{t}}\right)+\eta \operatorname{Gini}\left(\boldsymbol{w_{t}}\right)
    \end{equation}
\subsection{PG Risk Control Agent}
\begin{itemize}
    \item [1)]\textbf{State} Space\\
The state of the PG agent at time t is the same as that of the DDPG agent:
\begin{equation}
\boldsymbol{s}_{\boldsymbol{t}}=\boldsymbol{r}_{\boldsymbol{t}-T: \boldsymbol{t}}
\end{equation}
    
    \item [2)]\textbf{Action} Space\\
    The PG agent predicts the price movement after touching QPL, which can be divided into a bullish signal $S+$ and a bearish signal $S-$. The policy of PG agent at time t in our model is defined as:
\begin{equation}
\boldsymbol{p}_{\boldsymbol{t}}=
\left(p_{bullish},p_{bearish}\right)
\end{equation}
where each element in $\boldsymbol{p}_{\boldsymbol{t}}$  represents the probability of each type of action.\\
In the training environment, the action generated by the Monte Carlo method is given by:
\begin{equation}
    a_t = sampling(\boldsymbol{p_t})
\end{equation}
In the testing environment, the action selected with the largest probability is given by:
\begin{equation}
    a_t = argmax(\boldsymbol{p_t})
\end{equation}
\item[3)]QPLs-inspired Risk Control Mechanism\\
QPLs can play the role of a strong resistance and support line in the stock market, and the QPLs-inspired risk control mechanism proposed in this paper is based on intraday price fluctuations, in some cases the mechanism can be triggered to add to a position at a lower price or reduce it at a higher price to gain more profit.\\
Fig. \ref{fig7} shows a total of four types of intraday price movements that may trigger the early execution strategy.
%%%%%%%%%%
%%%%%%%%%%
% \begin{figure}[h]
%     \centering
%     \subfigure[]{\includegraphics[width=0.24\textwidth]{image/QPLKLINE--.png}} 
%     \subfigure[]{\includegraphics[width=0.24\textwidth]{image/QPLKLINE--.png}} 
%     \subfigure[]{\includegraphics[width=0.24\textwidth]{image/QPLKLINE--.png}}
%     \subfigure[]{\includegraphics[width=0.24\textwidth]{image/QPLKLINE--.png}}
%     \caption{(a) blah (b) blah (c) blah (d) blah}
%     \label{fig:foobar}
% \end{figure}
\begin{figure}[ht] 
  \begin{subfigure}[b]{0.5\linewidth}
    \centering
    \includegraphics[width=.95\linewidth]{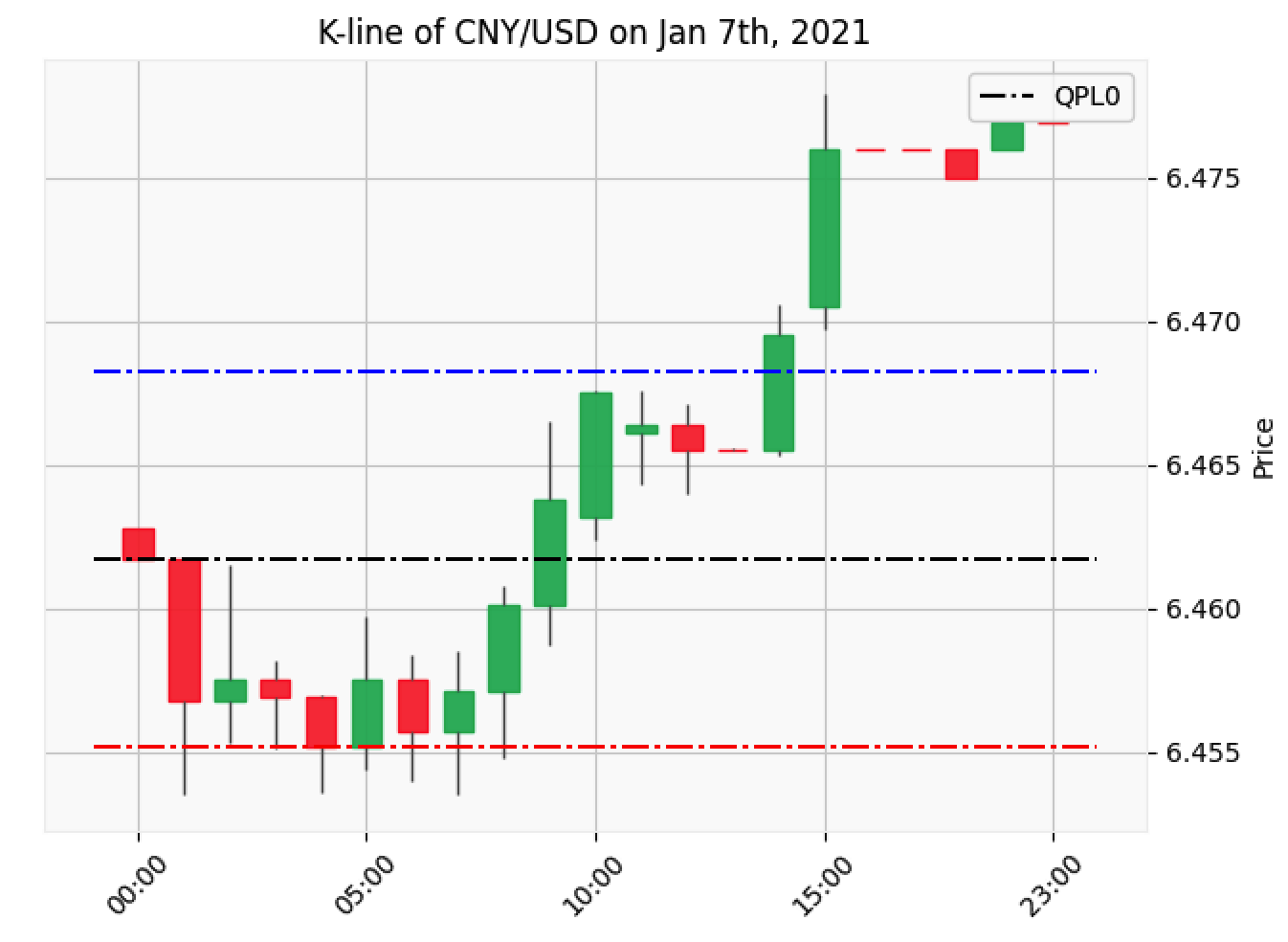} 
    \caption{$v_{i,t}^{low}<QPL^{-1}_{i,t}$, $S+$} 
    \label{fig7:a} 
    \vspace{4ex}
  \end{subfigure}%% 
  \begin{subfigure}[b]{0.5\linewidth}
    \centering
    \includegraphics[width=.95\linewidth]{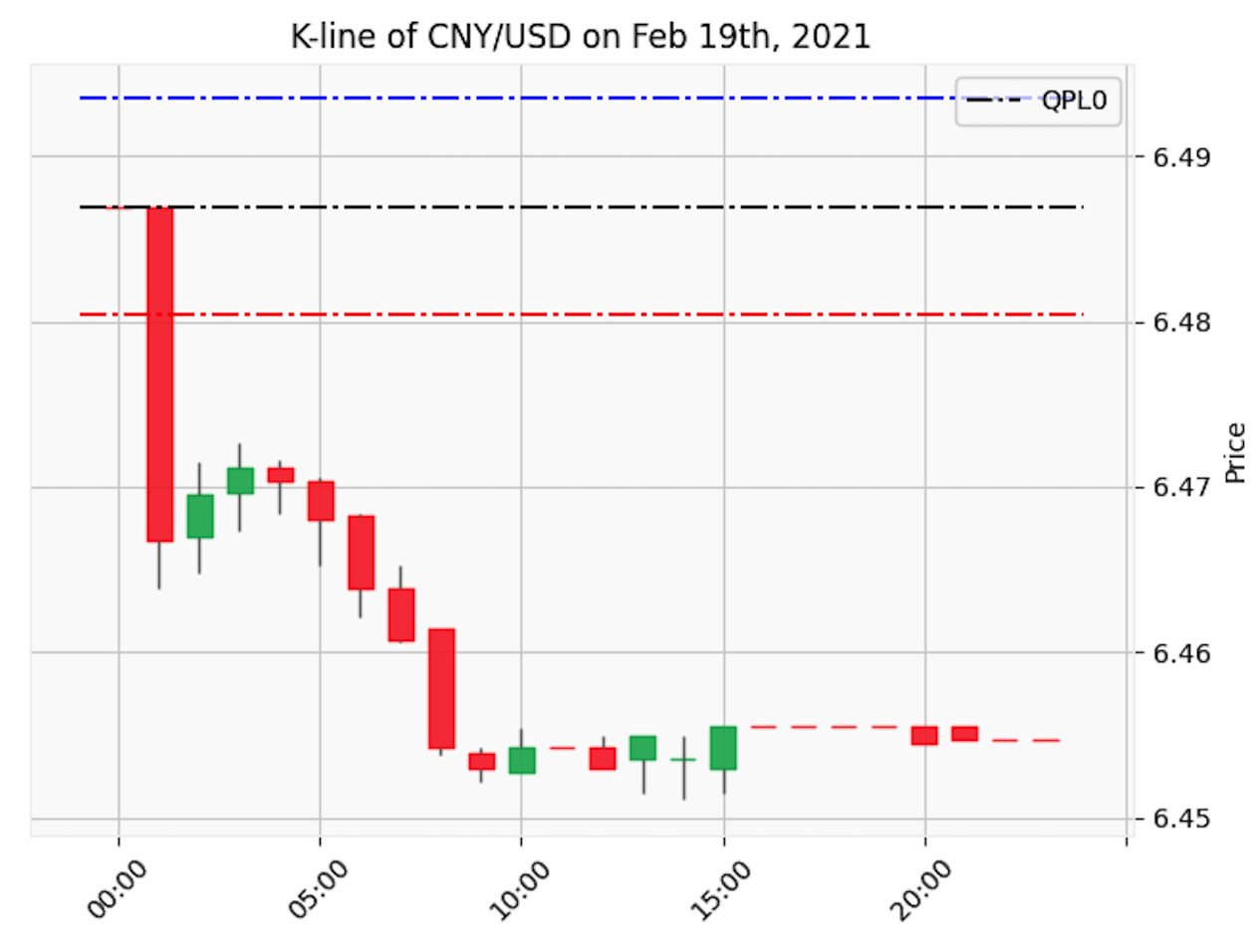} 
    \caption{$v_{i,t}^{low}<QPL^{-1}_{i,t}$, $S-$} 
    \label{fig7:b} 
    \vspace{4ex}
  \end{subfigure} 
  \begin{subfigure}[b]{0.5\linewidth}
    \centering
    \includegraphics[width=.95\linewidth]{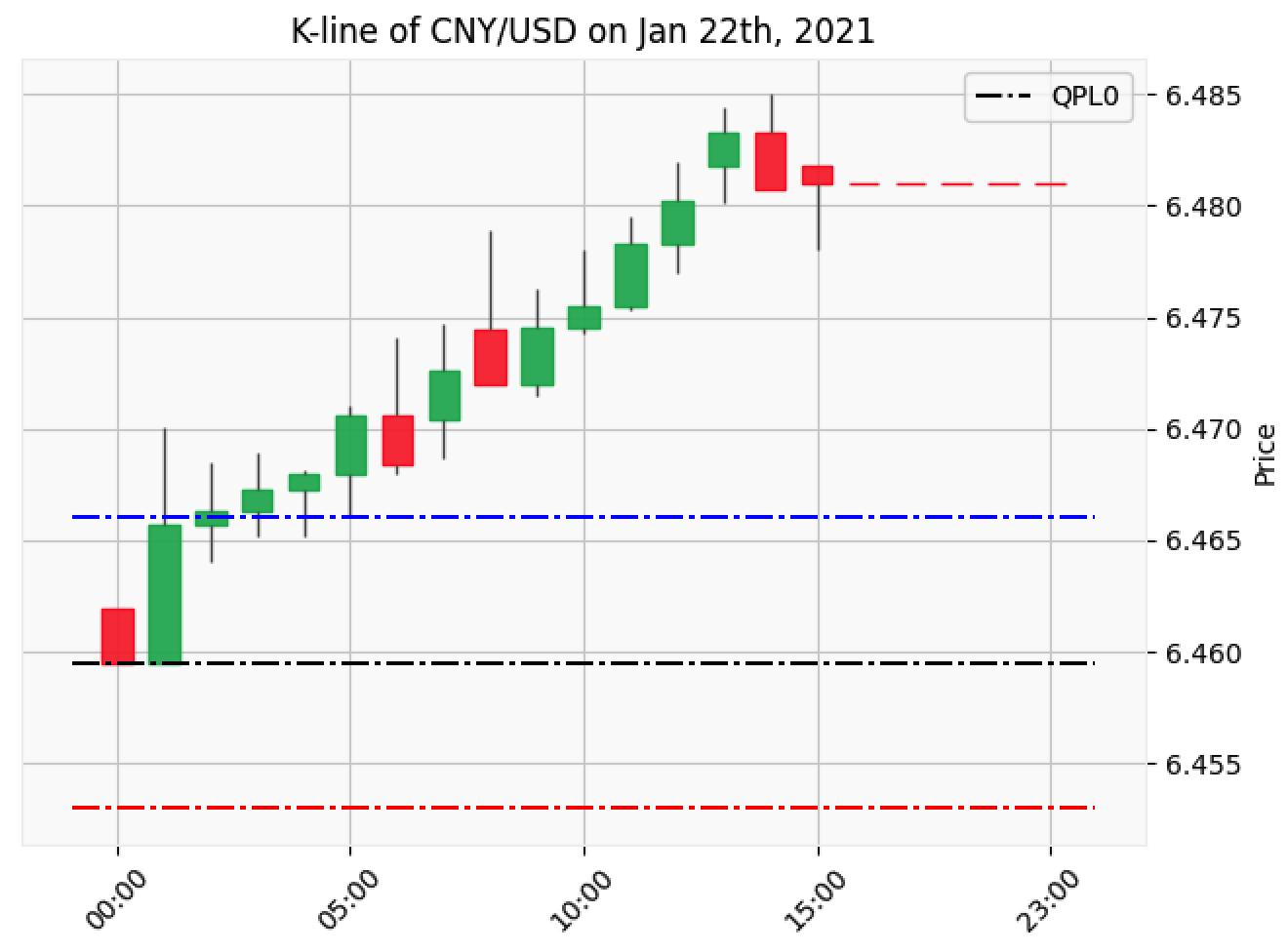} 
    \caption{$v_{i,t}^{high}>QPL^{+1}_{i,t}$, $S+$} 
    \label{fig7:c} 
  \end{subfigure}%%
  \begin{subfigure}[b]{0.5\linewidth}
    \centering
    \includegraphics[width=.95\linewidth]{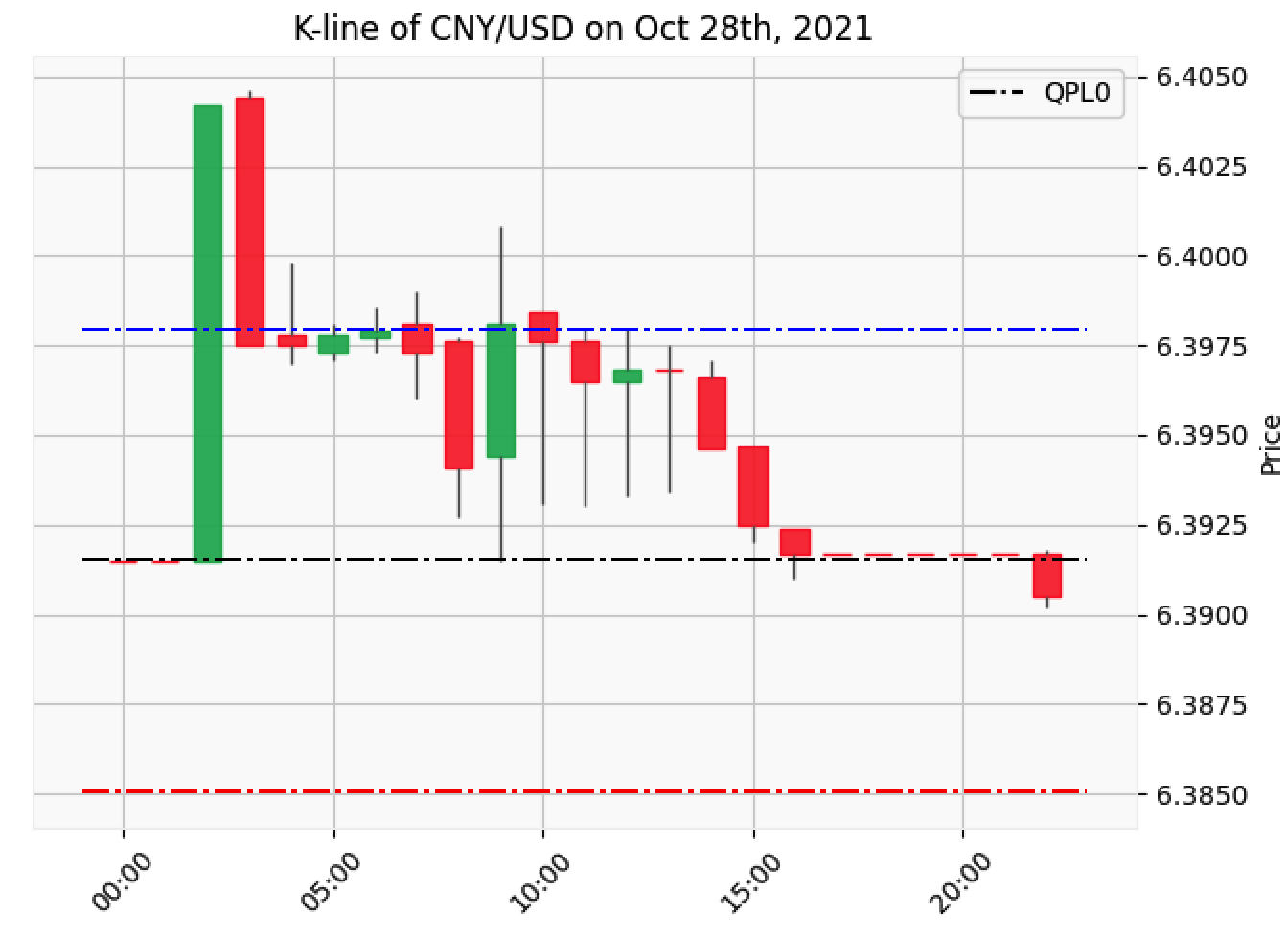} 
    \caption{$v_{i,t}^{high}>QPL^{+1}_{i,t}$, $S-$} 
    \label{fig7:d} 
  \end{subfigure} 
  \caption{Illustration of Various Price Movements}
  \label{fig7} 
\end{figure}

The detailed illustration of all the conditions is as follows, 
\begin{itemize}
    \item[a)] In the first condition (Fig.1 (a)\&(b)), the intraday price dips below $QPL^{-1}$, which is represented on the daily trading data as $v_{i,t}^{low}<QPL^{-1}_{i,t}$. In such case, if we receive a bullish signal $S+$ and the trading agent recommends buying $\Delta w_{i,t}>0 $, or we receive a bearish signal $S-$ and the trading agent recommends selling $\Delta\boldsymbol{w_{i,t}}<0 $, we should execute the trade early and set the execution price $v_{i,t}^E =  QPL^{-1}_{i,t}$.
    \item[b)] In the second condition (Fig.1 (c)\&(d)), the intraday price rises above $QPL^{+1}$ and does not dip to $QPL^{-1}$, which is represented by $v^{low}_{i,t}>QPL^{-1}_{i,t}$ and $v^{high}_{i,t}>QPL^{+1}$ on the daily trading data, in which case if we receive a bullish signal $S+$ and the trading agent recommends to buy $\Delta w_{i,t}>0 $, or we received a bearish signal $S-$ and the trading agent recommends selling $\Delta w_{i,t}<0 $, we should execute the trade early and set the execution price $v_{i,t}^E =  QPL^{+1}_{i,t}$.
    \item[c)]In other conditions, we are unable to trigger the risk control mechanism, and set the execution price $v_{i,t}^E$ as the closing price of the day $v^{close}_{i,t}$.
\end{itemize}
\item[4)] \textbf{Reward}\\
The reward of the PG agent is used to represent the return generated by the action it selected, so we can define it as:
\begin{equation}
    reward_t = \Delta \boldsymbol{w_t}\times(\frac{\boldsymbol{v_t^{close}} }{\boldsymbol{v_t^{E}}}-1)
\end{equation}
\end{itemize}
% Policy is directly modified in PG. Define a trajectory as $\tau=s_{1}, a_{1}, s_{2}, \ldots$ then the probability of each trajectory may be determined as follows:
% \begin{equation}
% \begin{aligned}
%     p_{\theta}(\tau)&=p\left(s_{1}\right) p_{\theta}\left(a \mid s_{1}\right) p\left(s_{2} \mid s_{1}, a_{1}\right) p_{\theta}\left(a_{2} \mid s_{2}\right) \ldots\\
%     &=p\left(s_{1}\right) \prod_{t=1}^{T} p_{\theta}\left(a_{t} \mid s_{t}\right) p\left(s_{t+1} \mid s_{t}, a_{t}\right)
% \end{aligned}
% \end{equation}

% Even when a precise action is taken, the state returned by the environment is still random. As a result, in PG, the aim is to maximize the cumulative anticipated reward, which is denoted as:
% \begin{equation}
%     \overline{R_{\theta}}=\sum_{\tau} R(\tau) p_{\theta}(\tau)=E_{\tau \sim p_{\theta}(\tau)}[R(\tau)]
% \end{equation}

\section{Implementation}
\subsection{System Architecture}
\begin{figure*}[h]
    % \centring
    \includegraphics[width=1\textwidth]{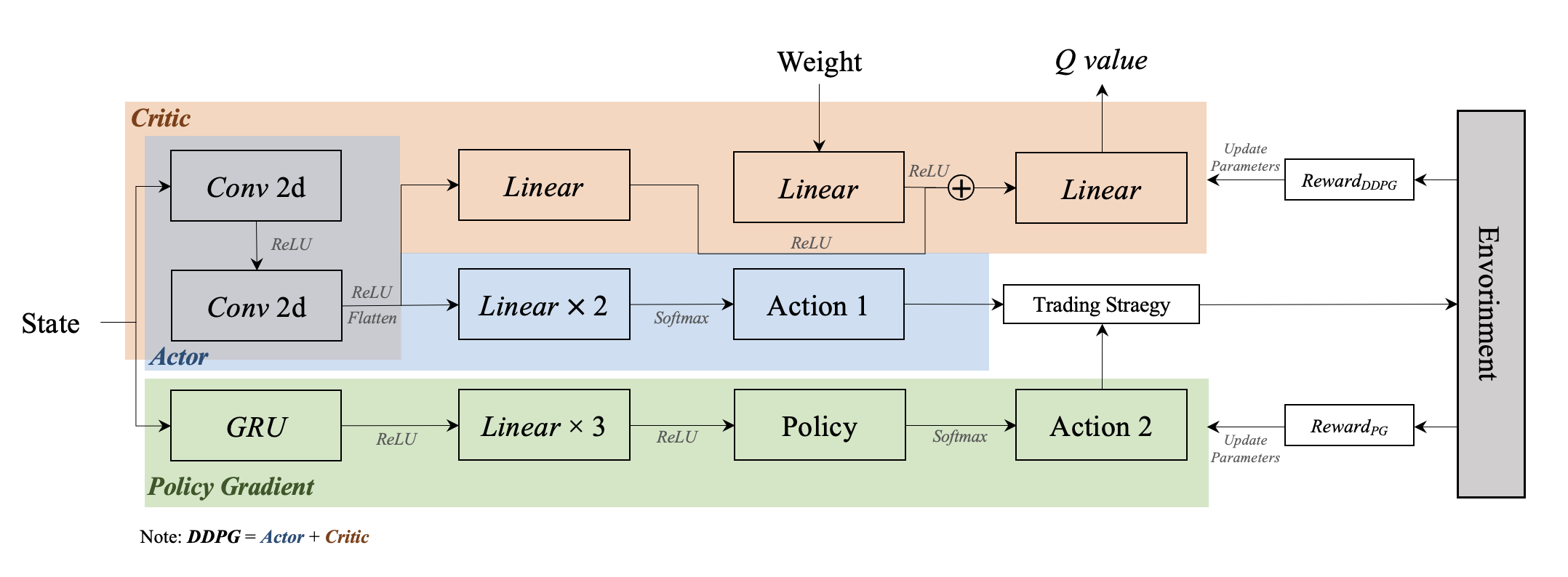}
    \caption{System Architecture}
    \label{fig:architecture}
\end{figure*}

Our Augmented DDPG model consists of two basic DRL models, Deep Deterministic Policy Gradient (DDPG) and Policy Gradient (PG) models, where the algorithm is described in Algorithm\ref{alg:algorithm1}. A schematic of our System is shown in Fig. \ref{fig:architecture}.
% \begin{document} 
\begin{algorithm}
	\caption{Augmented Deep deterministic Policy Gradient Algorithm} 
    \label{alg:algorithm1}
	\begin{algorithmic}[1]
            \State Randomly initialize online network $F(s,a|\theta)$ with actor $F_\mu(s|\theta_\mu)$, critic $F_Q(s, a|\theta_Q)$ and policy agent $F_\pi(s|\theta_\pi)$
            \State Initialize target network $F'$ with parameter $\theta' \leftarrow \theta$
            \State Initialize Experience Replay Buffer $\mathcal{R}$
		\For {episode = 1 to $M$}
                \State Initialize a random process $\mathcal{N}$ for action selection
                \State Receive initial observation state $s_1$
                \State Initialize policy reward list $\mathcal{L}$
			\For {t = 1 to T}
                    \State Get action $a_t = F_\mu(s_t|\theta_\mu)+\mathcal{N}_t$
                    \State Get policy $p_t = F_\pi(s_t|\theta_\pi)$
                    \State Execute action $a_t$ and policy $p_t$
                    \State Get reward $r_t$, policy reward $r_t^p$ and next state $s_{t+1}$
                    \State Save the policy reward $r_t^p$ into $\mathcal{L}$
				\State Save the transition $(s_t,a_t,r_t,s_{t+1})$ into $\mathcal{R}$
                    \State Sample N transitions $(s_i,a_i,r_i,s_{i+1})$ from $\mathcal{R}$
                    \State Set $y_i = r_i +\gamma F_Q'(s_{i+1},F_\mu'(s_{i+1}|\theta_\mu')|\theta_Q')$
                    \State Update the critic by minimizing the loss $$L = \frac{1}{N} \sum_{i}\left(y_{i}-F_Q\left(s_{i},a_{i} |\theta_{Q}\right)\right)^{2}$$
                    \State Update actor by sampled policy gradient 
                    $$\nabla_{\theta_{\mu}} J \approx \frac{1}{N} \sum_{i} \nabla_{a} F_Q\left(s_i, a | \theta_{Q}\right)\nabla_{\theta_\mu} F_\mu\left(s |\theta_{\mu}\right) | _{a = F_\mu(s_i)} $$
                    \State Softly update target network parameter 
                    $$\theta' \leftarrow \tau\theta+(1-\tau)\theta'$$
			\EndFor
                \State Calculate the cumulative reward $\psi_{t}$ for each policy reward in $\mathcal{L}$
                $$\psi_{t} = \sum_{t^{\prime}=t}^{T} \gamma^{t^{\prime}-t} r_{t^{\prime}}$$
                    
			\State Update policy agent parameter $$\theta_\pi=\theta_\pi+\alpha \sum_{t}^{T} \psi_{t} \nabla_{\theta_\pi} \log F_{\pi}\left(a_{t} \mid s_{t}\right)$$
		\EndFor
	\end{algorithmic} 
\end{algorithm}

% \end{document}
% \begin{figure}[H]
%     \centering
%     \includegraphics[width = 0.5\textwidth]{image/pesudo.png}
%     \caption{Pseudo Code for the Algorithm}
%     \label{fig:pseudo}
% \end{figure}

%%%%%%%%%%%%%%%%%%
%%%%%%%%%%%%%%%%%%
Based on previous work\cite{gao2020application} which concluded that in DPO task, using CNN as state encoder outperforms RNN such as LSTM in many cases when constructing actors and critics network of DDPG algorithm. As shown in the Fig. \ref{fig:architecture}, two layers of CNN are used as their State Encoder in building the networks of Actor and Critic, and unlike the traditional DDPG with separate Actor and Critic networks, our model shares one State Encoder for both Actor and Critic, the Integrated Actor-Critic structure will speed up the learning process and reduce sample complexity \cite{9923640}. After the state is encoded, it is passed into the Actor and Critic decoders respectively. In the Actor decoder, there are two fully connected layers and one softmax layer, and finally the action (weight) is output. In the Critic decoder, the CNN is followed by a fully-connected layer, and a dedicated fully-connected layer encodes the action (weight). The output of the state and the output of the action are combined in the fully-connected layer to produce the Q-value. For the Policy Gradient agent, since Gate Recurrent Unit (GRU) network is robust in handling sequential data\cite{cho2014learning}, we choose GRU as its state encoder for time series feature extraction, which is then output to two fully-connected layers and followed by a Softmax layer. The output is the probability of the policy, and finally a policy is randomly selected as the action (signal).

% \subsection{Environment Implementation}

% The actor in DDPG triggers an action $a_t$ to the environment. Each element in the action represents the value share of different products, where cash is the first element in the vector. PG will input a signal to the environment, signal is a set of one-dimensional vectors, with shape (1,m). Each element has two return values, one is the \textbf{bullish} signal \textbf{S+} and the other is the \textbf{bearish} signal \textbf{S-}. The bullish signal is triggered when the price hits a resistance or support line, and the price will continue to rise. And the bearish signal means that when the price hits a resistance or support line, the price will continue to stay down. Each element of the signal represents what PG considers to be the bullish or bearish condition of the stock.

% \subsection{QPLs-based Trading Strategy}
% As shown in figure 2, QPLs that reflected the intrinsic energy levels of finance products can be viewed as the support and resistance lines. Our trading strategy is based on such nature of QPLs. 
% \begin{figure}[H]
%     \centering
%     \includegraphics[width = 0.3\textwidth]{image/qpl.png}
%     \caption{QPL}
%     \label{fig:my_label}
% \end{figure}

\section{Experiments}
\subsection{Experiment Settings}
In our experiment, we selected five popular Forex products to comprise our portfolio: CNYUSD, EURGBP, EURJPY, EURUSD, and GBPJPY. The datasets for the five products were from Yahoo Finance, a widely used financial data platform. Each dataset has 2048 trading days, from February 2014 to December 2021. The first 80\% of the data was set for training our model and the rest 20\% was used as a test. In the back-testing, the initial portfolio value $p_0 = 100,000$, and the trading commission was 0.1\%. The size of the sliding window of the state space was set to $n=3$.

In Augmented DDPG, the network was trained for 50 episodes using the Adam optimizer. For the DDPG agent, the learning rates of the actor and critic networks were $1e-4$ and $1e-3$, respectively. For the target network, its soft learning rate was set to $1e-2$. For the neural network in the PG agent, the learning rate was set to $1e-4$.

\subsection{Baselines Introduction}
To demonstrate the superiority of our model, we choose the following baseline models to compare with ours,
\begin{itemize}
    \item[1)] \textbf{Original DDPG}\\
    The Original DDPG\cite{https://doi.org/10.48550/arxiv.1509.02971} model is chosen to demonstrate a significant improvement in our Augmented DDPG.

    \item[2)] \textbf{Uniform Constant Rebalanced Portfolios (UCRP)}\\
    UCRP \cite{maclean2011kelly} maintains the same wealth allocation among a collection of assets day to day.

    \item[3)] \textbf{Online Newton Step (ONS)}\\
    ONS \cite{10.1145/1143844.1143846} is a portfolio management algorithm that is based on the Newton Method and makes use of the functions' second derivative.

    \item[4)] \textbf{Winner}\\
    Follow the Winner \cite{followWinner} is a trading strategy that always puts its attention on the asset that is performing well and shifts all of the portfolio weights there.

    \item[5)] \textbf{The Best Asset}\\
    The Best Asset is the asset that has the highest accumulated return through the testing period.
\end{itemize}

\subsection{Metrics Introduction}
For evaluation, we make use of the following performance metrics: Annual Rate of Return (ARR), Sharpe Ratio (SR), and Maximum Drawdown (MDD).

Detailed illustrations of these metrics are as follows,
\begin{itemize}
    \item[1)] \textbf{Annualized Rate of Return (ARR)}\\
    Annualized Rate of Return is determined as the equivalent annual return received by an investor over a specified time period \cite{Annualized}, which is formulated as,
    \begin{equation}
    A R R=\frac{V_{f}-V_{i}}{V_{i}} \times \frac{T_{\text {year }}}{T_{\text {all }}}
    \end{equation}
    where $V_i$ and $V_f$ are the initial value of the portfolio and the final value of the portfolio respectively; $T_{year}$ and $T_{all}$ represent the number of trading days in a year and the total number of trading days respectively.

    \item[2)] \textbf{Annualized Sharpe Ratio (ASR)}\\
    The Sharpe Ratio divides a portfolio's excess returns by a measure of its volatility to assess risk-adjusted performance \cite{Sharpe}. It also represents a measure of the additional return an investor obtains for every unit of increased risk. SR is given by,
    \begin{equation}
    ASR=\frac{\bar{R}}{\sigma_{R}}
    \end{equation}
    where $\bar{R}$ is the annualized excess return, and $\sigma_R$ represents the standard deviation of the portfolio’s annualized excess return.

    \item[3)] \textbf{Maximum Drawdown (MDD)}\\
    The largest observed loss in a portfolio from its peak to its trough until a new peak is reached is known as the Maximum Drawdown. Over a given time frame, Maximum Drawdown serves as a predictor of downside risk \cite{mdd}, which is given by,
    \begin{equation}
    M D D=\max _{\tau \geq t} \frac{p_{t}-p_{\tau}}{p_{t}}
    \end{equation}
    where $p_t$ represents the peak value and $p_\tau$ represents the trough value.
\end{itemize}

\subsection{Results and Discussion}
\subsubsection{Back-testing Result}

\begin{figure}[!h]
    \centering
    \includegraphics[width = 0.48\textwidth]{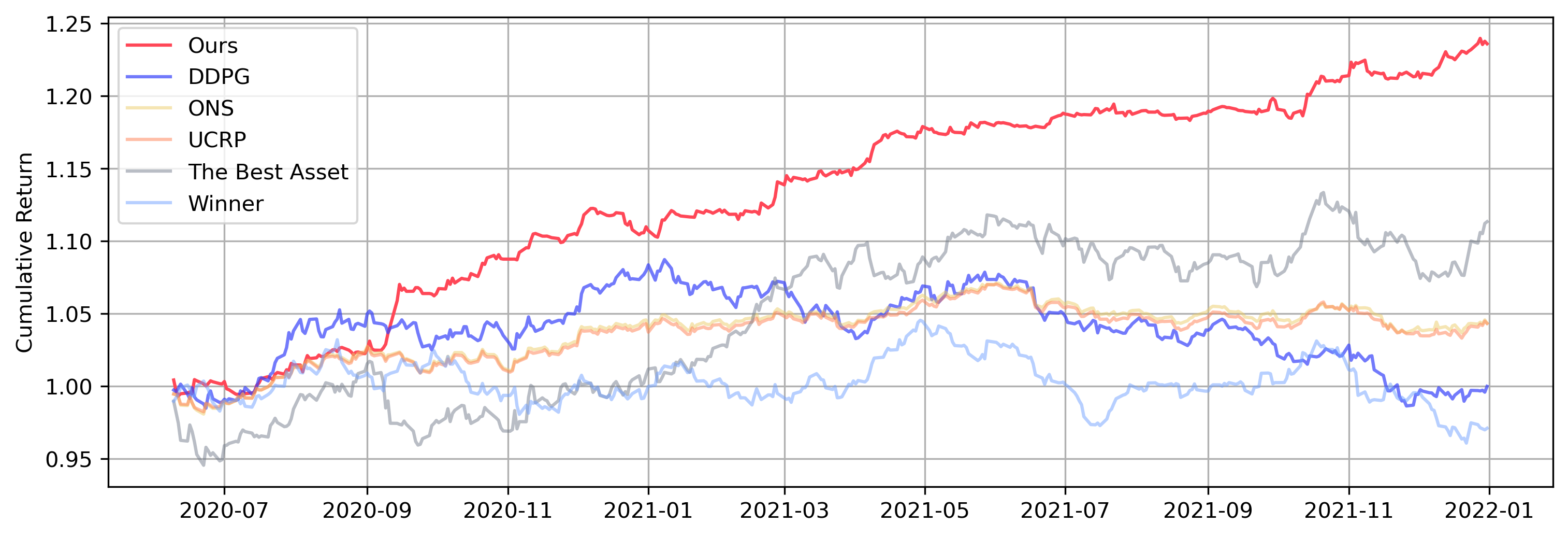}
    \caption{Cumulative Returns of Different Models}
    \label{fig:backtest}
\end{figure}

\begin{figure}[!h]
    \centering
    \includegraphics[width = .48\textwidth]{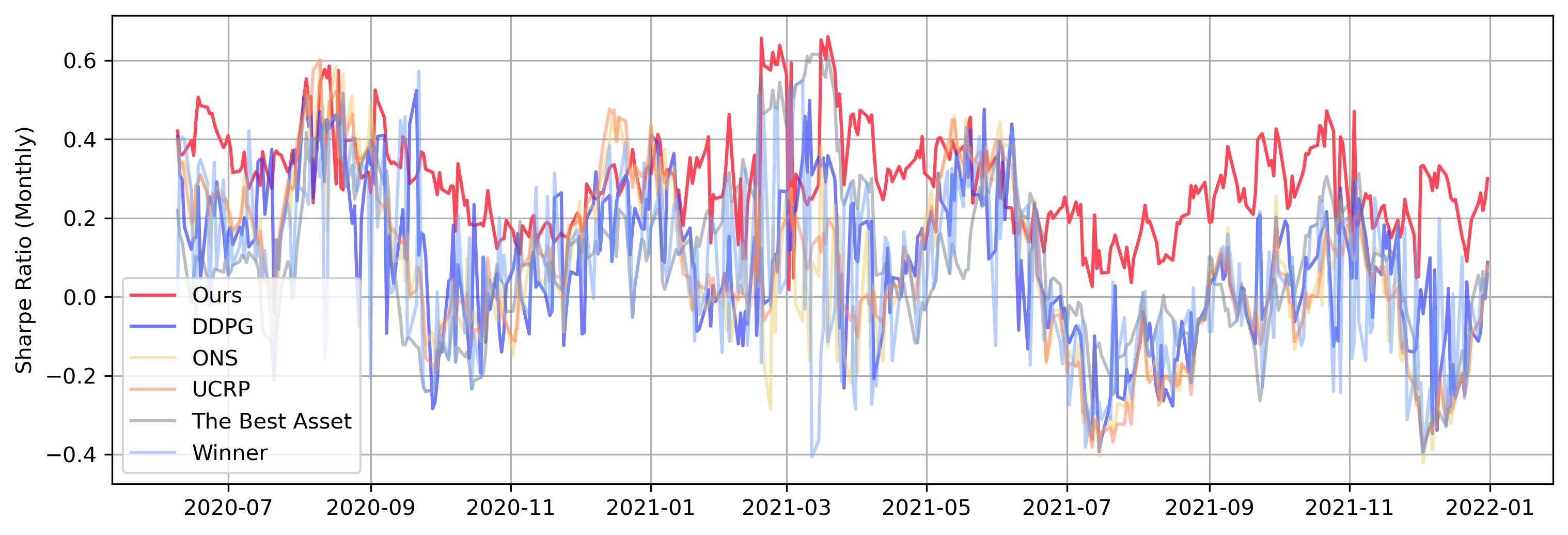}
    \caption{Sharpe Ratio of Different Models}
    \label{fig:sr}
\end{figure}

In Fig. \ref{fig:backtest} we can see that the market for the period from July 2020 to January 2022, where the test set is located, is generally a bull market. Even though the other baseline models were able to make some profit in the market, it is clear to see that our model has greater profitability as compared with these benchmark models. Our model has a significant advantage over the original DDPG. And within these models, only our model outperforms the Best Asset. In terms of risk control, we can see from TABLE I that our model achieves not only the highest cumulative return, but also the lowest maximum drawdown, which proves that our model is well-controlled in terms of risk while pursuing profits. Also, we can see from Fig. \ref{fig:sr} that our model keeps the portfolio at a relatively ideal Sharpe ratio throughout the test time period without excessive volatility like other baselines, which proves that our model's risk control is significantly better than other models.

\begin{table}[!h]
\centering
\begin{tabular}{cccc}
\hline
Models & MDD & ARR & ASR \\ \hline
Ours & \textbf{1.7695\%} & \textbf{1.2104} & \textbf{1.0811} \\
DDPG & 9.2718\% & 0.9999 & 0.0196 \\
ONS & 3.1938\% & 1.0387 & 0.3129 \\
UCRP & 3.4794\% & 1.0386 & 0.3151 \\
Winner & 8.0903\% & 0.9742 & -0.0555 \\
The Best Asset & 5.6688\% & 1.1011 & 0.3398 \\ \hline
\end{tabular}
\label{table:1}
\caption{Performance Metrics of Different Models}
\end{table}

\subsubsection{Analysis of Augmented DDPG}
\begin{figure}[!h]
    \centering
    \includegraphics[width = 0.48\textwidth]{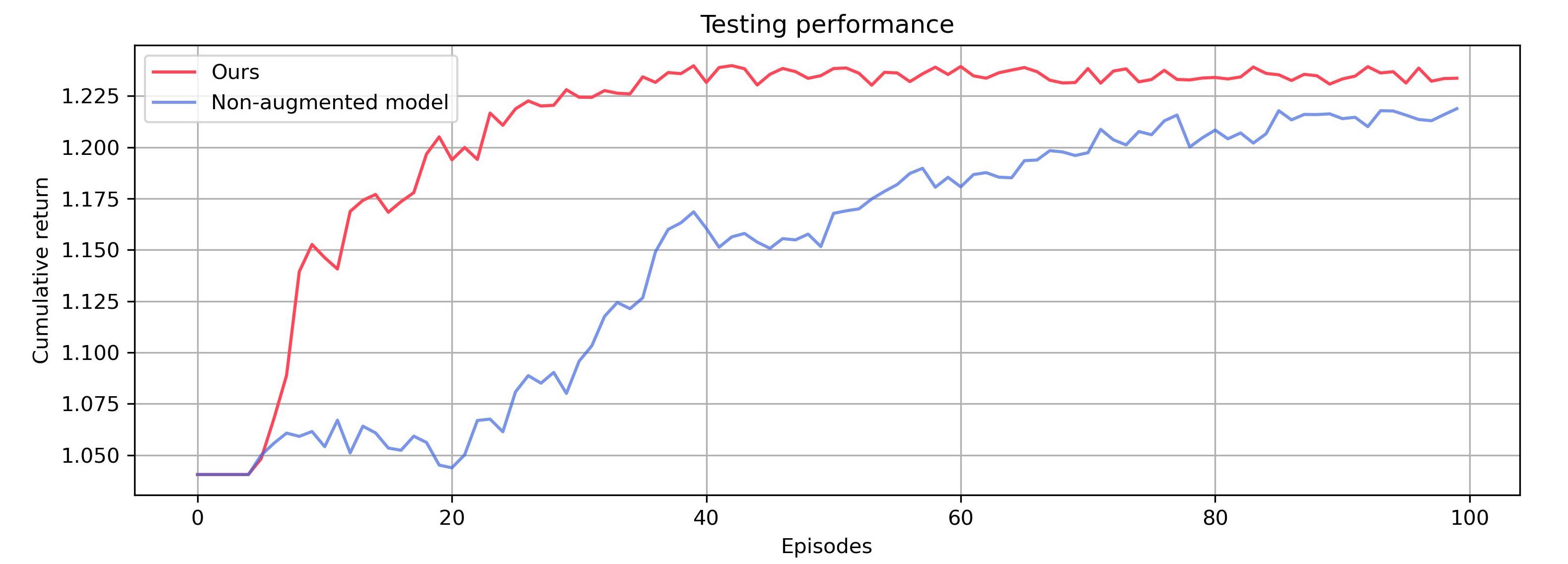}
    \caption{Comparison of Training Efficiency}
    \label{fig:testPer}
\end{figure}

As we have mentioned, our model Augmented DDPG is proposed by borrowing ideas from the DDPG agent with an integrated Actor-Critic structure and the PG agent. Therefore, our ablation analysis took our model to compare to the DDPG agent with a separate Actor-Critic structure and the PG agent, which we also call the non-augmented model. As illustrated in Fig. \ref{fig:testPer}, our model achieves high returns in the test set faster, provided the same number of episodes are trained, which demonstrates that our changes to the model's intrinsic structure significantly improve the efficiency of training and reduce the sample complexity.

%As illustrated in Fig. \ref{fig:actorloss}, the convergence speed of the Actor Loss function of our model is significantly improved compared to the non-augmented model. Fig. \ref{fig:testPer} also shows another aspect of the improvement brought by our model, which is that our model achieves high returns in the test set faster, provided the same number of episodes are trained.

% \begin{figure}[htbp]
% \begin{minipage}[t]{0.5\linewidth}
% \centering
% \includegraphics[height=4cm,width=4cm]{image/actorloss.png}
% \caption{Actor Loss}
% \label{fig:actorloss}
% \end{minipage}%
% \begin{minipage}[t]{0.5\linewidth}
% \centering
% \includegraphics[height=4cm,width=4cm]{image/testingPer.png}
% \caption{Performance}
% \label{fig:testPer}
% \end{minipage}
% \end{figure}

\subsubsection{Analysis of Gini Bonus}
\begin{figure}[htbp]
    \centering
    \includegraphics[width = .48\textwidth]{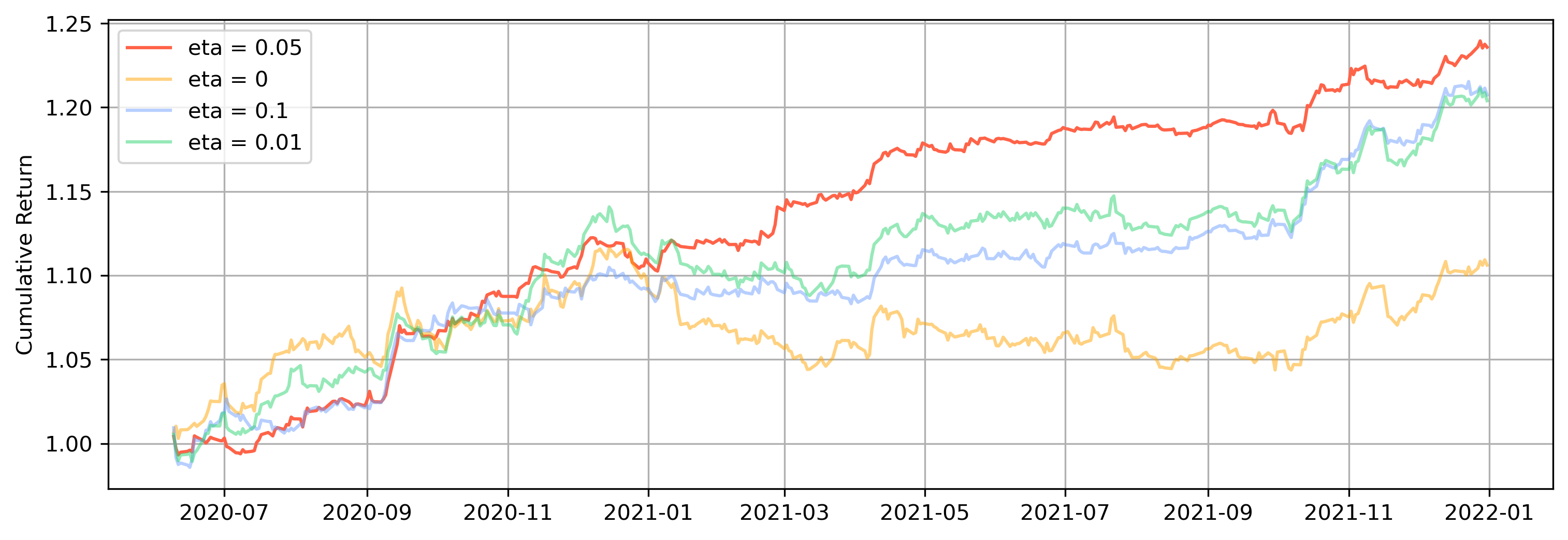}
    \caption{Cumulative Returns with Different Gini Bonus Coefficients}
    \label{fig:ginitest}
\end{figure}

As we expect, the weight given by Augmented DDPG will be more diverse when the coefficient becomes larger. When the coefficient becomes smaller, Augmented DDPG will be biased toward investing in a single asset. However, through experiments, we found that we can find a balance in which the model better balances profit and risk. As shown in TABLE II, when $\eta=0.05$, the model can achieve the highest return and Sharpe ratio, but the lowest MDD is achieved when $\eta$ is set to 0.1. Figure \ref{fig:ginitest} also gives an intuitive illustration of our findings.

\begin{table}[htbp]
\centering
\begin{tabular}{cccc}
\hline
\multicolumn{1}{l}{} & \textbf{MDD} & \textbf{ARR} & \textbf{ASR} \\ \hline
$\eta=0$ & 6.5062\% & 1.0947 & 0.7533 \\
$\eta=0.01$ & 4.6247\% & 1.1820 & 0.6801 \\
$\eta=0.05$ & 1.7695\% & \textbf{1.2104} & \textbf{1.0811} \\
$\eta=0.1$ & \textbf{1.7422\%} & 1.1851 & 0.7533 \\ \hline
\end{tabular}
\caption{Performance Metrics with Different Gini Bonus Coefficients}
\end{table}

\section{Conclusion and Future Work}
In this paper, we proposed and implemented a multi-agent portfolio trading system based on the Augmented DDPG and quantum finance-based trading strategies. The experiment result reveals that our model improves the training efficiency by sharing the encoder network of Actor and Critic networks. Also, by realizing the interaction between the trading decision agent and the risk control agent, the balance between profit and risk in forex investment is found in our system.

Through introducing the Gini bonus of the position distribution to the reward function, we make the decision agent prefers to output more balanced positions. 
%Also, the QPL-inspired trading strategy controlled by the PG agent minimizes potential risk by anticipating the price trend after touching QPL.
Also, the QPLs-inspired trading strategy controlled by the PG agent improves the model's ability to react to potential profits and risks by anticipating the price trend after touching QPLs.

In experiments, by conducting back-testing on the 5 popular forexes’ historical data, our system outperforms all baseline models that are popular in the DPO field by 20\%+ in terms of Annualized Rate of Return. This demonstrates our trading system is of significance in the pursuit of high returns and risk control in the DPO field.

To further enhance the risk management capabilities of our trading system, our future work will combine our trading system with the NLP models to capture news and information that can influence the forex market.

\section*{Acknowledgment}
% The authors would like to thank the anonymous reviewers
% for their constructive suggestions and comments on our paper.
% This work was supported in part by the Guangdong Provincial Key
% Laboratory of Interdisciplinary Research and Application for Data
% Science, BNU-HKBU United International College, project code
% 2022B1212010006, and in part by Guangdong Higher Education
% Upgrading Plan (2021-2025) with UIC research grant R0400001-22.
% Meanwhile, this work was supported by Research Grant R202008 of 
% BNU-HKBU United International College
% and Guangdong Province F1 project grant on Curriculum Development and Teaching Enhancement on Quantum Finance course UICR0400050-21CTL.
This paper was supported by Research Grant R202008 of Beijing Normal University-Hong Kong Baptist University United International College (UIC), Key Laboratory for Artificial Intelligence and Multi-Model Data Processing of Department of Education of Guangdong Province, Guangdong Province F1 project grant on Curriculum Development and Teaching Enhancement on Quantum Finance course UICR0400050-21CTL and by the Guangdong Provincial Key Laboratory of Interdisciplinary Research and Application for Data Science, BNU-HKBU United International College (2022B1212010006).

\bibliographystyle{IEEEtran}
\bibliography{IEEEabrv,mybibfile}
\end{document}